\documentclass[%
 rsi,
 amsmath,amssymb,
 reprint,%
 groupedaddress,%
 superscriptaddress,
]{revtex4-1}

\usepackage{graphicx}
\usepackage{dcolumn}
\usepackage{bm}
\usepackage{algorithmic}
\usepackage{textcomp}
\usepackage{booktabs}
\usepackage{amsfonts,dsfont}
\usepackage{xcolor} 
\usepackage[caption=false,font=footnotesize]{subfig}
\usepackage[T1]{fontenc}
\usepackage{mathptmx}
\begin{document}

\preprint{AIP/123-QED}

\title[]{Invariant Representations in Deep Learning for Optoacoustic Imaging}

%

\author{M. Vera}
\affiliation{Universidad de Buenos Aires, Facultad de Ingenier\'ia, Paseo Col\'on 850, C1063ACV, Buenos Aires, Argentina.}%
\affiliation{Consejo Nacional de Investigaciones Cient\'ificas y T\'ecnicas, (CONICET), Godoy Cruz 2290, C1425FQB, Buenos Aires, Argentina.}%
\author{M. G. Gonz\'alez}%
\email{The author to whom correspondence may be addressed: mggonza@fi.uba.ar}%
\affiliation{Universidad de Buenos Aires, Facultad de Ingenier\'ia, Paseo Col\'on 850, C1063ACV, Buenos Aires, Argentina.}%
\affiliation{Consejo Nacional de Investigaciones Cient\'ificas y T\'ecnicas, (CONICET), Godoy Cruz 2290, C1425FQB, Buenos Aires, Argentina.}%
\author{L. Rey Vega}%
\affiliation{Universidad de Buenos Aires, Facultad de Ingenier\'ia, Paseo Col\'on 850, C1063ACV, Buenos Aires, Argentina.}%
\affiliation{Consejo Nacional de Investigaciones Cient\'ificas y T\'ecnicas, (CONICET), Godoy Cruz 2290, C1425FQB, Buenos Aires, Argentina.}%


\begin{abstract}
Image reconstruction in optoacoustic tomography (OAT) is a trending learning task highly dependent on measured physical magnitudes present at sensing time. The large number of different settings, and also the presence of uncertainties or partial knowledge of parameters, can lead to reconstructions algorithms that are specifically tailored and designed to a particular configuration which could not be the one that will be ultimately faced in a final practical situation. Being able to learn reconstruction algorithms that are robust to different environments (e.g. the different OAT image reconstruction settings) or \emph{invariant} to such environments is highly valuable because it allows to focus on what truly matters for the application at hand and discard what are considered spurious features. In this work we explore the use of deep learning algorithms based on learning invariant and robust representations for the OAT inverse problem. In particular, we consider the application of the ANDMask scheme due to its easy adaptation to the OAT problem. Numerical experiments are conducted showing that, when out-of-distribution generalization (against variations in parameters such as the location of the sensors) is imposed, there is no degradation of the performance and, in some cases, it is even possible to achieve improvements with respect to standard deep learning approaches where invariance robustness is not explicitly considered.
\end{abstract}

\maketitle

\section{Introduction}
\label{sec:introduction}

Optoacoustic tomography (OAT) is an imaging technique based on the optoacoustic (OA) effect. Through the use of an appropriate illumination source and ultrasonic detectors, OAT combines the high contrast imaging of optical techniques while keeping precise spatial resolution of ultrasonic detection \cite{xu2006}. When biological tissue is illuminated with non-ionizing short laser pulses, a rapid increase in temperature leads to the formation of pressure waves due to the thermoelastic expansion of the sample under study. Acoustic waves propagate through the sample and are finally sensed by wideband ultrasonic transducers, typically placed around the sample \cite{paltauf2017,Awasthi_Jain_Kalva_Pramanik_Yalavarthy_2020}. The detected signals (commonly known as the sinogram) are used in properly designed numerical algorithms to recover the initial pressure profile induced by the light absorption. As optical absorption is linked with several physiological properties, among others oxygen saturation and hemoglobin concentration, several diagnostic applications can be based on this technique \cite{Hauptmann_Cox_2020,tian2020}.

Deep learning is an emerging research area, in which specialized artificial neural networks are used for pattern recognition and machine learning tasks \cite{LeCun_Bengio_Hinton_2015}. In particular, convolutional neural networks (CNN) are widely used for different imaging tasks \cite{convnet_original} due to its excellent ability to generalize to unseen examples not present in the training dataset, but identically distributed to them. One of its greatest exponents today is the UNet architecture \cite{Ronneberger_Fischer_Brox_2015}, which is basically a multi-scale convolutional autoencoder using a residual connection between input and output and skip connections that connect encoder and decoder at each scale, providing, among other things, numerical stability during the backpropagation training. In OAT, this architecture has been widely used in image post-processing to reduce artifacts originating from limited view, sparse sampling or limited bandwidth \cite{antholzer2019, davoudi2019,godefroy2021}. Moreover, the different scales at the encoder and decoder could include dense connectivity, such as in the Fully-Dense UNet architecture (FD-UNet), allowing a better information flow through the network, robustness against learning redundant features and numerical stability. The use of FD-UNet in OAT is promising, delivering excellent restoration results and artifacts suppression \cite{guan2020}. 

Data collection in the OAT area presents challenges, especially from the difficulty and cost of obtaining properly labeled samples (that is, having a good number of ground-truth images to be reconstructed). It also requires a precise specification of the experimental imaging system and also a sophisticated modeling of the physics of the problem with an accurate knowledge of a number of parameters, like the speed of sound, which heavily depends on the sample under study. It is useful to think that each experimental imaging setting, with some of its relevant parameters such as sensor positions, angular coverage of detection curve, etc, defines  an \emph{environment} and that the exact distribution of the data collected is conditional to such an environment. As the specific details of the specific environment considered for the training phase of the image reconstruction algorithm cannot be exactly the same as in the testing phase (because of the presence of small changes or uncertainties in the OAT system setting), a good practice could be to look for a learning procedure that is invariant to such environment particularities. A robust concept of generalization should consider, not only the capability of performing well on unseen examples with the same distribution to those observed during training (good generalization for a specific environment) but also to be able to perform well on examples from other environments not necessarily available during training.

The assumption that data pairs sinogram and ground-truth image will be independent and identically distributed (IID) relative to the application-collected data, it is clearly an oversimplification of the experimental conditions. This standard and well-known concept of IID generalization between equally distributed datasets (at training and testing) is then clearly insufficient and the need to a better definition of generalization arises. This concept of harder and broader generalization is known as horizontal, strong, or out-of-distribution (OOD) generalization \cite{Scholkopfetal21}. Basically, an algorithm has well OOD generalization capacity when it is limited to learning characteristics that are invariant to reasonable changes in the environment and that are important for the specific problem at hand. When a machine learning algorithm spends part of its capacity in learning particular environment details which are not crucial for the task, we say that \emph{spurious learning} is taking place. An example of spurious learning in the literature is the task of deciding the presence of cows or camels in images \cite{Beery18}. Most learning algorithms tend to misclassify when they come across images of cows in deserts or camels in grasslands. In the OAT context, a spurious learning situation would arise when some capacity of the model would be spent to learn (during the training phase) information about the experimental setup (e.g. the location of the ultrasonic sensors), that would not exactly match the scenario during the testing phase.

One of the first considered schemes specifically designed to cope with this problem is the  \emph{Invariant Causal Prediction} (ICP) \cite{petters16} algorithm, which only applies to linear models and scales exponentially with respect to the number of variables in the learning problem. Another approach is the \emph{Invariant Risk Minimization} (IRM) \cite{Arjovsky19} proposal which is a learning paradigm that attemps to generate representations invariant to multiple training environments. Another important invariant learning algorithm is the Inter-environmental Gradient Alignment (IGA) \cite{Koyama20}, which minimizes the error on each environment while reducing the variance of the gradient of the loss per environment. Finally, the ANDMask method \cite{parascandolo21}  minimizes the error on the different environments used at training by updating the model in those directions where the sign of the gradient of the loss is the same for most environments. 
In this paper, we explore the use of the ANDMask approach in the image reconstruction problem for OAT, looking for a robust reconstruction method against variations or partial knowledge of different physical magnitudes involved in the problem. The proposal of using the ANDMask algorithm is mainly due to the facts that this method achieves the best performance in the \emph{linear unit-tests for invariance discovery} \cite{Aubin21}, it is extremely simple to implement with a negligible computational overhead with respect to standard backpropagation procedure and it can be easily adapted to denoising or inverse problems such as image reconstruction in OAT. 

The paper is organized as follows. In Section \ref{sec:reconstruction} we introduce the  acoustic reconstruction problem in OAT. In Section \ref{sec:ood} we cover the basics of the ODD generalization concept and the ANDMask algorithm. The implementation details of the simulations are presented in Section \ref{sec:matmet}. Then, in Section \ref{sec:experimentos}, the merits of our proposal are evaluated numerically. Finally, in Section \ref{sec:conclu}, some concluding remarks are discussed.

\section{The reconstruction problem in OAT}
\label{sec:reconstruction}

\subsection{The forward problem}
\label{subsec:met_forward}

It is well-known that  after the excitation of a biological sample by an electromagnetic pulse $\delta(t)$, the acoustic pressure $p(\mathbf{r},t)$ at position $\mathbf{r} \in\mathbb{R}^3$ and time $t$, satisfies \cite{Wang_Wu_2007}:

\begin{equation}
\left(\frac{\partial^2}{\partial t^2} - v_s^2 \, \nabla^2 \right) p(\mathbf{r},t) = 0
\label{eq:waveeq}
\end{equation}

\noindent with the initial conditions,

\begin{equation}
p(\mathbf{r},0) = p_0(\mathbf{r})\, \text{,} \quad \left(\partial p /\partial t\right)(\mathbf{r},0)= 0 
\label{eq:waveeq_cond_ini}
\end{equation}

\noindent where $p_0(\mathbf{r})$ is the initial OA pressure and $v_s$ represents the speed of sound in the medium, which is assumed to be lossless and homogeneous. Under the usual hypothesis of thermal and acoustic confinement \cite{Kruger_Liu_Fang_Appledorn_1995}, that is, when the laser pulse duration is short enough such that the heat conduction and acoustic propagation into neighboring regions of the illuminated region can be neglected, the initially induced pressure $p_0(\mathbf{r})$ is proportional to the total absorbed optical energy density. In this sense, $p_0(\mathbf{r})$ is also usually referred as the image. Using Green's function formalism, the pressure received by an ideal point-detector at position $\mathbf{r_d}$ can be written as \cite{xu2006}:

\begin{equation}
p_d(\mathbf{r_d},t)=\frac{1}{4\pi\,v_s^2} \frac{\partial}{\partial t}\iiint_{V} \, p_0(\mathbf{r}) \frac{\delta\left(t-|\mathbf{r_d}-\mathbf{r}|/v_s\right)}{|\mathbf{r_d}-\mathbf{r}|} d\mathbf{r}
\label{eq:fo_time}
\end{equation}

The goal of the OAT inverse problem is to reconstruct $p_0(\mathbf{r})$ from the signals $p_d(\mathbf{r_d},t)$ measured at various positions $\mathbf{r_d}$ {(collectively known as the sinogram)}, which are typically in a surface that contains the volume of interest \cite{lutzwieler2013}.

\subsection{The inverse problem}
\label{subsec:met_inverse}
Possibly, the most popular reconstruction approach in OAT are back-projection (BP) type algorithms, due to their simple implementation and applicability in several practical imaging scenarios \cite{rosenthal2013}. One of the most important formulations of the BP approach is the universal back-projection algorithm \cite{xu2005}, which presents a closed form solution to the problem of image reconstruction that can be exactly formulated for several popular geometries (e.g., spherical, cylindrical or rectangular). Although extremely popular, BP methods presents some drawbacks. They assume point detectors with isotropic angular response and no bandwidth limitations  \cite{rosenthal2013}. However, in practice, the transducers are extended, have a limited bandwidth and their spatial response is anisotropic. For these non-ideal imaging scenarios, the BP method could significantly deviate from reality, generating imaging artifacts and distorted images. Moreover, BP reconstruction formula assumes that the detected signals are not noisy, which does not occur in practice.

\begin{figure}
    \centering
	\includegraphics[width=0.95\columnwidth]{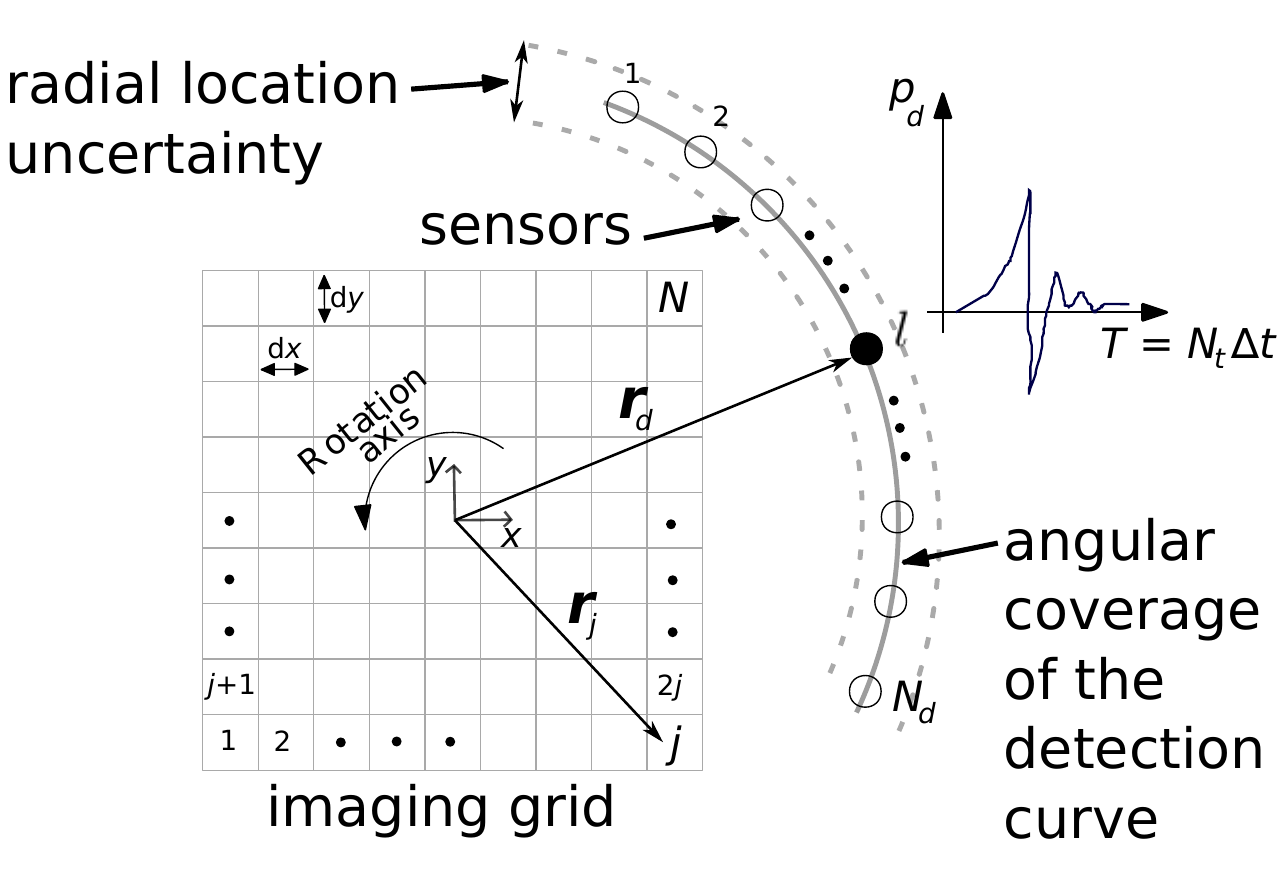}
	\caption{Schematic of the OA imaging setup used in the simulations. Sensor are uniformly distributed in a circle around the sample. The angular coverage refers to the angular section in the circle where the sensors are located. For example, a $360^{\circ}$ angular coverage means that the sensors cover the whole circle, while a $90^{\circ}$ angular coverage means that the sensors cover a quarter of the circle.}
	\label{fig:setup}
\end{figure}

A different approach to the reconstruction problem is given by a model-based-matrix (MBM) formulation \cite{Rosenthal_Razansky_Ntziachristos_2010}. In this case, the forward solution in \eqref{eq:fo_time} is discretized. As a result a matrix equation is obtained, which is used for solving the inverse problem. One of the advantages of this approach is that any linear effect in the system may be easily considered (e.g. sensor form factors, linear filtering or the spatial response of the sensors). The linear system can be written as: 

\begin{equation}
\mathbf{p_d}=\mathbf{A} \, \mathbf{p_0}
\label{eq:mbt}
\end{equation}

\noindent where the sinogram $\mathbf{p_d} \in\mathbb{R}^{N_d \cdot N_t\times 1}$ is a column vector representing the measured pressures at a set of detector locations $\mathbf{r_d}_l$ ($l=1 \ldots N_d$) and time instants $t_k$ ($k=1 \ldots N_t$); $\mathbf{p_0} \in\mathbb{R}^{N\times 1}$ is a column vector representing the values of the initial acoustic pressure on the imaging region grid; and $\mathbf{A} \in\mathbb{R}^{N_d \cdot N_t\times N}$ is the model matrix. The $j$-th element ($j=1 \ldots N$) in $\mathbf{p_0}$ contains the average value of the initial pressure within a volume element of size $\Delta V$ at position $\mathbf{r}_j$. Once the discrete formulation has been established, the inverse problem is reduced to the algebraic problem of finding a solution to the linear system in \eqref{eq:mbt}. Notice that this system is not neccesarily invertible. In this way, a different number of techniques, as variants of least squares problems can be used \cite{rosenthal2013},\cite{Hirsch_Gonzalez_ReyVega_2021}. The matrix $\mathbf{A}$ can be written as the multiplication of two matrices $\mathbf{A^{oa} \, A^s}$ where $\mathbf{A^s}$ represents the response function of the imaging system for an ideal point-like sensor and $\mathbf{A^{oa}}$ is the matrix form of a time derivative operator. The matrix $\mathbf{A^s}$ is defined as \cite{paltauf2018, dean2012, wang2011}: 

\begin{equation}
    A^s_{lkj} = \frac{1}{4\pi v_s^2}\frac{\Delta V}{\Delta t^2} \frac{d(t_k,\mathbf{r}_j,\mathbf{r}_{dl})}{|\mathbf{r_d}_l - \mathbf{r}_j|}
\label{eq:Gs1}
\end{equation}

\begin{equation}
    d(t_k,\mathbf{r}_j,\mathbf{r}_{dl}) = \begin{cases}
    1 & \text{if  }\, |t_k - \frac{|\mathbf{r_d}_l - \mathbf{r}_j|}{v_s}| < \Delta t/2  \\
    0 & \text{otherwise} 
  \end{cases}
    \label{eq:Gs2}
\end{equation}

\noindent where $\Delta t$ is the time step at which the signals $p_d(\mathbf{r_d},t)$ are sampled. In the case of a ﬁnite-size detector, the spatial impulse response (SIR) of the sensor is taken into account by dividing the area of the sensor into surface elements (treated as point detectors) which are then added up \cite{paltauf2018}. The OAT imaging setup studied in this work is shown in Fig. \ref{fig:setup}. We chose this type of two-dimensional OAT system because it was demonstrated to be very useful for proof-of-concept studies due to its simplicity, low cost, and effectiveness \cite{tian2020}. 

Finally, one of the most basic method for image reconstruction is given by \cite{Awasthi_Jain_Kalva_Pramanik_Yalavarthy_2020}:

\begin{equation}
\mathbf{p_{lbp}}=\mathbf{A}^T \, \mathbf{p_d}
\label{eq:lbp}
\end{equation}

\noindent where $\mathbf{p_{lbp}} \in\mathbb{R}^{N\times 1}$ is the reconstructed image written as a vector and $\mathbf{A}^T$ is the transpose of $\mathbf{A}$. This approach is known as Linear Back-Projection (LBP) because $\mathbf{A}^T$ behaves as a backprojection operator that maps from pressure data to the image space \cite{Hauptmann_Cox_2020}. Although the quality of the image delivered by this method is not always really good (specially in constrained image systems with, for example, a low number of sensors), it can be easily implemented with a reasonable computational effort. It is for this reason that, when deep learning is applied to OAT, the LBP method is commonly used to obtain an initial image that is fed to the neural network to be improved.

\subsection{Deep Learning in OAT image reconstruction}

The current paradigm in OAT image reconstruction is, in essence, a denoising autoencoder \cite{Goodfellow_Bengio_Courville_2016}. From the original biological tissue $Y$, and through laser illumination the sinogram $S$ is generated, which is transformed into a corresponding image $X$ via LBP method. As the quality of the phantom $X$ recovered by LBP could be insufficient and introduce artifacts, for example,  when the number of sensor is not sufficiently large or when the signal to noise ratio (SNR) is not high enough, a deep neural network $f_\theta(\cdot)$ can be fed with this image to deliver a reconstructed image $\hat{Y}=f_\theta(X)$ with better quality and a reduced number of artifacts. The parameters $\theta$ of the network have to be properly optimized using a training dataset and a meaningful loss function for the application at hand. Although different neural networks architectures can be considered, in this paper we would restrict ourselves to the FD-UNet architecture because of its excellent performance in OAT image reconstruction \cite{guan2020} as was mentioned in the introduction. 

\section{OOD-Generalization and the ANDMask algorithm}
\label{sec:ood}

Consider the case in which at training time there are a number of different datasets that are not identically distributed. As explained in the introduction this could be due to the fact that the datasets are conditional to different \emph{environments}. In the context of OAT, different environments correspond to typical different experimental conditions. For example, uncertainties in sensor positions, variations in the speed of sound of the samples under study, different detection curves, etc.

Following the framework introduced in \cite{petters16} and improved in \cite{Arjovsky19}, we consider datasets $\mathcal{D}_e = \{(X^e_i, Y_i^e)\}_{i=1}^{n_e}$ collected under multiple training environments $e\in\mathcal{E}_{\text{tr}}$, where $\mathcal{E}_{\text{tr}}$ is considered discrete and finite. The dataset $\mathcal{D}_e$, from environment $e$, contains $n_e$ examples identically and independently distributed according to some probability distribution conditioned on environment $e$. These environments include the pair of random variables $(X, Y)$ measured under the conditions specified by the corresponding environments. In the OAT context $X^e_i$ is the $i$-th measured sinogram of the environment $e$ and $Y^e_i$ is the corresponding true image. Then, our goal is to use these multiple datasets to learn the parameters of neural network $f_\theta$, such that the reconstructed images $\hat{Y}=f_\theta(X)$ performs well across a large set of unseen but related environments $\mathcal{E}_{\text{all}}$ which obviously contains the environments $\mathcal{E}_{\text{tr}}$ observed during training. Namely, we wish to minimize the OOD-Risk:

\begin{equation}\label{eq:oodrisk}
R^\text{OOD}(f_\theta)= \max_{e\in\mathcal{E}_{\text{all}}} R^e(f_\theta)    
\end{equation}

\noindent where $R^e(f_\theta)=\mathbb{E}[\ell(f_\theta(X^e),Y^e)]$ is the IID-risk under environment $e$ and $\ell(\cdot,\cdot)$ is the loss function. Here, the set of all environments $\mathcal{E}_{\text{all}}$ contains all possible experimental settings. The goal of OOD generalization problem is to train an algorithm that generalizes well on all environments, but using only a small subset of them at training time. Popular approaches in this field use the hypothesis that such a predictor shall be an invariant predictor that captures the essential aspects between inputs $X$ and outputs $Y$ that are common to the different environments.

Two sources of errors are defined for the learning algorithms (considering that there are no bias issues in neural architectures): lack of data and spurious learning (that is, variability between environments) which are related with IID and OOD generalization, respectively. 
On the one hand, having a small dataset could mask the error due to spurious learning, as the main issue here is the lack of data to properly train the deep network to have good generalization capabilities even when the testing data is exactly distributed as the training data. On the other hand, the spurious learning issue is not solved with more data (it would still be present with infinite data \cite{Arjovsky19}). Therefore, when a large dataset is available, spurious learning  would be the only relevant source of error. In this context, huge datasets would be more profitable for invariant algorithms, if they are correctly designed.

There is an important number of algorithms based on invariant representations in the bibliography. As previously mentioned, in this work we decided to implement the ANDMask algorithm \cite{parascandolo21} due to its performance \cite{Aubin21} and to the fact that it can be easily adapted to the OAT problem. This is because this method does not assume any additional hypothesis over the existence of an invariant distribution, it simply searches for a predictor that is as invariant as possible based on a specific design guideline of its training process. In its most simple implementation, ANDMask zero out those gradient components with respect to weights that have inconsistent signs across environments. Formally, the  gradients $\nabla_\theta R^e(f_\theta)$ for each environment $e$ and used at each iteration of the backpropagation algorithm are masked using $m_\tau\odot \left(\frac{1}{d}\sum_{e=1}^d\nabla_\theta R^e(f_\theta)\right)$, where $\odot$ denotes Hadamard product and $m_\tau$ is a vector that vanishes the less common components $[m_\tau]_j$ of $m_\tau$ with

\begin{equation}
[m_\tau]_j=\mathds{1}\left\{\tau d\leq\left|\sum_{e=1}^d\text{sign}([\nabla_\theta R^e(f_\theta)]_j)\right|\right\}.    
\end{equation}

\noindent Note that the number of training environments $d$ limits the number of $\tau$ values for which the algorithm will show significant changes in its performance. For example, if $d=5$, the coincidence or not in the sign of the gradients can be 5-0, 4-1 or 3-2. The $\tau$ associated with these values will be 1, 0.6 and 0.2, respectively.

\section{Implementation Details}
\label{sec:matmet}
In this work, we numerically explore the merits of the ANDMask algorithm in the OAT image reconstruction task against variations in the number of sensors, angular coverage of the detection curve, and position and sound speed uncertainties. Sound speed and position uncertainties clearly introduce a difference between the data generation model and the construction of matrix $\mathbf{A}^T$ of the LBP stage and are not under control by the system designer. In that sense, these situations can be easily considered under the framework of different environments detailed above. See \cite{Hirsch_Gonzalez_ReyVega_2021, Tick_Pulkkinen_Tarvainen_2019, Sahlstrom2020, deanben2017} for the importance of considering the uncertainties in the sensor positions and the speed of sound. On the other hand, the number of sensors and the angular coverage of the detection curve are magnitudes directly associated with the experimental setup and are design choices. It is clear that, when those design choices are made, they are final and they are representative of a single environment from which all measurement will come. However, there are situations (e.g. the malfunction of a sensor or a group of them) that are not aligned with the assumption of a unique working environment used for both for the training phase and testing phase. Moreover, the problem of the variation of the angular coverage of the detection curve is important in its own, as it is related with the issue of the important practical situation of limited view OAT, where the detection curve over which the sensor are positioned has not a full angular coverage of the sample \cite{Schwab2019,davoudi2019,liu2013}. It is for these reasons that we included the analysis of different environments corresponding to variations in the number of sensors and angular coverage of detection curve.


Fig. \ref{fig:setup} shows the role of above mentioned magnitudes in the imaging setting and represents a two-dimensional OAT system implemented with $N_d$ sensors (32 is the nominal value) around the imaging region where the sample, uniformly illuminated, is placed. A square imaging region having a size of $7.2 \text{ mm} \times 7.2 \text{ mm}$ and a resolution of $128 \times 128$ pixels (pixel size of $dx = dy = 60\, \mu\text{m}$) was used and the sensors (circles) were placed equidistantly on an arc of circumference of radius $r_d = 10 \text{ mm}$. Although large area detectors are typically used in practice, for simplicity, we assumed the sensor to be point-like with a bandwidth of 15 MHz. For data collection, the time step $\Delta t$ was 10 ns with $N_t=$ 1024 samples. The nominal value of the speed of sound was set to $v_s= 1490 \text{ m/s}$ and the medium was assumed homogeneous with no absorption or dispersion of sound. The transducer frequency response was modeled using a band-pass filter with upper and lower cutoff frequencies $0.1 \text{ MHz}$ and $15 \text{ MHz}$, respectively. Moreover, white Gaussian noise was added with variable variance levels leading to sinogram measurement SNR values between $40 \text{ dB}$ and $60 \text{ dB}$. It is important to remark that, we choose a nominal small value (32) of $N_d$ because a low number of simultaneously acquired signals reduces the complexity and associated costs of OAT imaging systems \cite{davoudi2019}.

In order to train the network, we considered an initial number of $360$ images (experimental retinal vasculature phantoms obtained from a public database \cite{drive2020}) that were segregated into $7$ environments: $5$ for training and $2$ for testing purposes. For each environment, data augmentation was performed (basically rotations and translations) leading to $2216$ images each one. In addition, the different training sets were divided into $1776$ samples for the training procedure and $440$ for validation using early stopping. It is important to mention that, based on this number of images per environment, our simulations have a limited amount of data (something that is not unusual in an application such as OAT). Therefore, following the discussion in Section \ref{sec:ood} about lack of data and spurious learning, the ANDMask algorithm is in an unfavorable situation. 
For generating the corresponding sinograms that are used as input in the networks, we used (\ref{eq:Gs1}) considering the setup of Fig. \ref{fig:setup} which, according to our numerical simulations, gives very similar results to using the popular k-Wave toolbox \cite{kwave}, but requires significantly less computation time.

We trained the model during $100$ epochs with a batch-size of $2$ per training environment and a learning rate of $5\times10^{-4}$ with ADAM as optimizer \cite{Kingma2015AdamAM} and using the empirical mean square error (MSE) between the ground-truth images and the reconstructed ones as loss function.That is, for each environment $e$ included in the training dataset:
\begin{equation}
R^e(f_\theta)=\frac{1}{n_e}\sum_{i=1}^{n_e}\|\mathbf{p}_0^{i}-f_\theta(\mathbf{A}^T\mathbf{p}_d^i)\|^2,
    \label{eq:loss_emp}
\end{equation}
where $\mathbf{p}_0^i$ is the $i^{th}$ ground-truth initial pressure example for environment $e$ and $\mathbf{p}_d^i$ is the $i^{th}$ measured sinogram corresponding to $\mathbf{p}_0^i$. In this case matrix $\mathbf{A}$ is constructed with the nominal values of $\mathbf{r}_{\mathbf{d}}$ and $v_s$.

The network optimization was implemented in Python 3.8 with Pytorch and the equipment used for the simulations was a CPU Intel i7-9700F, 64 GB of RAM and a GPU RTX 2080 with 8 GB of memory. Testing environments were labeled as ``lax'' and ``challenging''. The lax environment did not exactly match the training environments but had features similar to them, while the challenging environment was characterized by parameters that fall outside the range used to generate the training environments. The so-called lax environment was used to test how well the trained model responds to environments that were not necessarily present in the training environments but that are somewhat close to them. On the other hand, the challenging environment was be used to evaluate how the model behaves for environments that deviate significantly from those used during training. That is, we analyzed different levels of OOD generalization for the trained model. Since we considered $5$ training environments, we used $\tau=0.4$ and $\tau=0.8$ which led us to implement ANDMask with 4 and 5 matching gradient signs respectively. 

We compared our results with a ``benchmark'' neural network whose model is identical but which is trained using a standard stochastic gradient descent scheme without the ANDMask algorithm and with a batch-size of $10$. In this respect, the ``benchmark'' is optimized using the same total number of training data but without being segregated according to the different environments. In addition, we reported the linear LBP results since this is the first and simple approximation to the true image that is fed as input of both, the ``benchmark'' neural network and the one trained using ANDMask. We evaluated the performance on the quality reconstruction with different popular metrics: Structural Similarity Index (SSIM), Pearson Correlation (PC), Root Mean Square Error (RMSE) and Peak Signal to Noise Ratio (PSNR).

\begin{figure*}
    \centering
    \includegraphics[width=2\columnwidth]{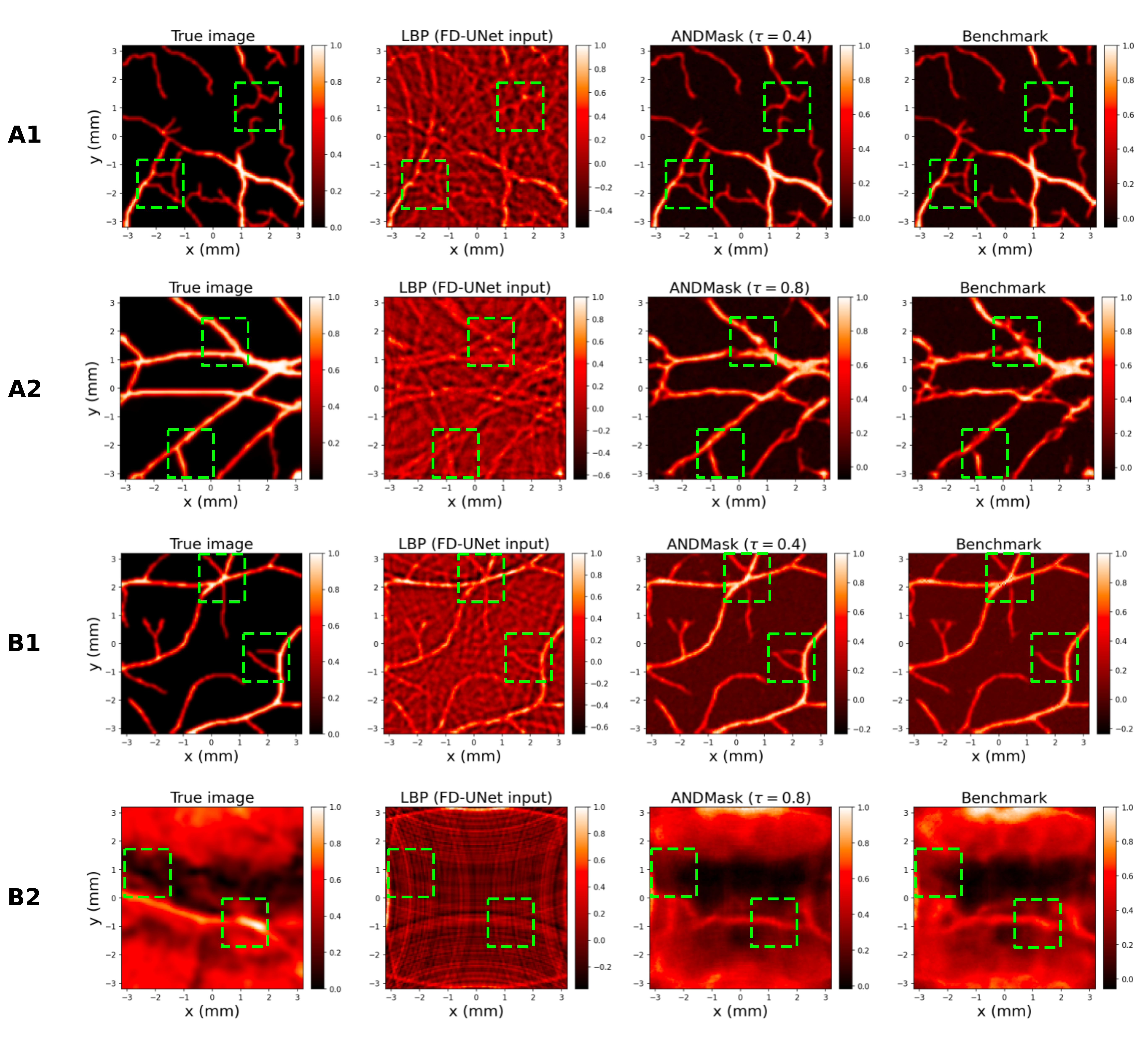}
    \caption{Examples of the reconstructed images obtained by the methods studied in this work. \textbf{A}: position uncertainty. \textbf{B}: number of detectors. \textbf{1}: Lax testing environment. \textbf{2}: Challenging testing environment.}
    \label{fig:results_1}
\end{figure*}

\begin{figure*}
    \centering
    \includegraphics[width=2\columnwidth]{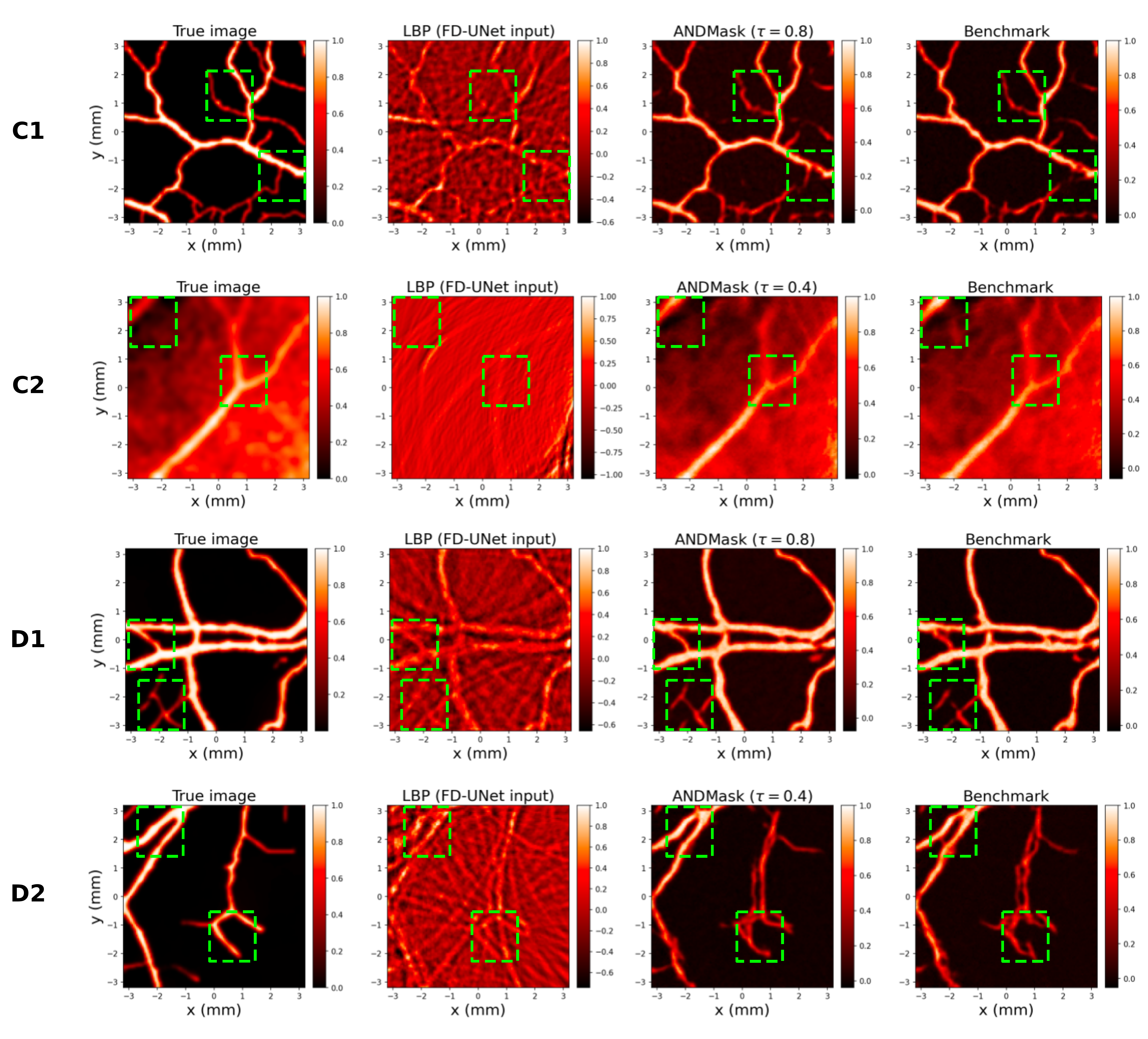}
    \caption{Examples of the reconstructed images obtained by the methods studied in this work. \textbf{C}: angular coverage of the detection curve. \textbf{D}: sound velocity value uncertainty. \textbf{1}: Lax testing environment. \textbf{2}: Challenging testing environment.}
    \label{fig:results_2}
\end{figure*}

\section{Numerical Results}
\label{sec:experimentos}

In Tabs. \ref{tab:position_uncertainties}-\ref{tab:vel_uncertainties} we show the average reconstruction performance, over 500 images not contained in the training database, of the methods LBP and the FD-UNet trained with and without ANDMask approach against variations in the detector position uncertainty (Tab. \ref{tab:position_uncertainties}), number of sensors (Tab. \ref{tab:num_detectors}), angular coverage of the detection curve (Tab. \ref{tab:arc}) and sound speed uncertainty (Tab. \ref{tab:vel_uncertainties}).

Figs. \ref{fig:results_1} and \ref{fig:results_2} show examples of the reconstructed images representing the qualitative results of Tabs. \ref{tab:position_uncertainties}-\ref{tab:vel_uncertainties}. It can be appreciated that the LBP reconstructions (second column of Figs. \ref{fig:results_1} and \ref{fig:results_2}) present serious artifacts and blurs caused by sparse sampling, limited view and uncertainty in the detectors locations and speed of sound values. Cases $A1$, $B1$ and $D1$ show the sparse sampling effect in which arc-shaped structures separated by the angle subtended by adjacent detector positions are clearly visible. The arc intensity diminishes when increasing the number of OA signals, as it can be seen in $B1$ ($N_d = 48$). The opposite case is presented in B2 ($N_d = 4$) where it is very difficult to recognize the shape of the original image. In $C1$ and $C2$, besides sparse sampling, there are also the limited view effect that produces streak image artifacts nearby the reconstructed location of absorbers (blood vessels). Moreover, when there is an appreciable uncertainty value in $r_d$ or $v_s$ ($> 1 \%$), such as cases B2 and D2, the images present blurring and motion effects. In all LBP reconstructions, there are negative values intensity that have no physical interpretation. As it is obvious, the artifacts mentioned above limit the spatial resolution and degrade image contrast.

In general, as shown in the third and forth columns (ANDMask and benchmark, respectively), the multi-scale architecture of the FD-UNet efficiently detect and remove streak-type artefacts and negative values from OA images reconstructed. In some cases, in agreement with the computed figure of merits values, the ANDMask approach gives better results than traditional training (benchmark). This is highlighted with green dashed-line squares in Figs. \ref{fig:results_1} and \ref{fig:results_2}. In the next subsections, we describe in detail each of the four cases studied in this work.

\subsection{Results for environments with different position uncertainties}

\begin{table}
\caption{Algorithm performance against different position uncertainties}
\label{tab:position_uncertainties}
\setlength{\tabcolsep}{3pt}
\begin{tabular}{p{45pt}|p{70pt}|p{25pt}|p{23pt}|p{23pt}|p{23pt}|}
\hline\hline
Environment&Algorithm & SSIM & PC & RMSE & PSNR\\
\hline
Lax&ANDMask $\tau=0.8$&0.873&0.968&0.048&27.375\\
Lax&ANDMask $\tau=0.4$&\bf{0.911}&\bf{0.981}&\bf{0.035}&\bf{30.241}\\
Lax&Benchmark&0.892&0.974&0.042&28.751\\
Lax&LBP&-0.125&0.231&0.563&5.620\\\hline\hline
Challenging&ANDMask $\tau=0.8$&\bf{0.692}&0.851&\bf{0.102}&\bf{20.457}\\
Challenging&ANDMask $\tau=0.4$&0.670&\bf{0.856}&0.113&19.946\\
Challenging&Benchmark&0.664&0.842&0.110&19.883\\
Challenging&LBP&0.036&0.244&0.413&8.001\\
\hline\hline
\end{tabular}
\end{table}

In this experiment, the number of detectors is $32$, the angular coverage in $360^{\circ}$ (sensors uniformly distributed) and the sound velocity is $1490 \text{ m/s}$ (without uncertainty). We define the training environments with sensor position uncertainties of $0.0001\,\%, 0.001\,\%, 0.01\,\%, 0.1\,\%$ and $1.0\,\%$, respectively. Regarding the testing environments, the lax environment is defined with a position uncertainties of $0\,\%$ and the challenging one with $1.5\,\%$.  The average results over both lax and challenging testing dataset are presented in Tab. \ref{tab:position_uncertainties}. We see that the ANDMask shows advantages over the benchmark neural network for all the observed metrics. In particular, $\tau=0.4$ (only discard gradients where 3 environments have one sign and the other 2 the opposite) is the best hyperparameter for the lax environment and $\tau=0.8$ (discard $3$-$2$ and $4$-$1$ cases) is the best for the challenging one. {In Fig. \ref{fig:results_1} (rows $A1$ and $A2$) some qualitative results are presented for both lax and challenging testing environments for the simple LBP reconstruction and the neural network trained with and without ANDMask. It can be seen that, the reconstructed images from ANDMask and benchmark have a good quality and almost all structures from the original image are recovered. However, the neural network trained with ANDMask preserves better some finer details (see green dashed-line boxes). This is specially true in the challenging environment where the blood vessels reconstructed by the benchmark network have some visible shape irregularities that are not present in the original image.}


\subsection{Results for environments with different number of detectors}

\begin{table}
\caption{Algorithm performance against different number of detectors}
\label{tab:num_detectors}
\setlength{\tabcolsep}{3pt}
\begin{tabular}{p{45pt}|p{70pt}|p{25pt}|p{23pt}|p{23pt}|p{23pt}|}
\hline\hline
Environment&Algorithm & SSIM & PC & RMSE & PSNR\\
\hline
Lax&ANDMask $\tau=0.8$&0.843&0.965&0.066&24.314\\
Lax&ANDMask $\tau=0.4$&0.831&0.967&0.060&\bf{25.053}\\
Lax&Benchmark&\bf{0.847}&\bf{0.968}&\bf{0.064}&24.845\\
Lax&LBP&-0.156&0.320&0.537&6.192\\\hline\hline
Challenging&ANDMask $\tau=0.8$&\bf{0.517}&0.519&0.203&14.111\\
Challenging&ANDMask $\tau=0.4$&0.489&0.531&\bf{0.202}&\bf{14.156}\\
Challenging&Benchmark&0.516&\bf{0.534}&\bf{0.202}&14.135\\
Challenging&LBP&0.054&0.222&0.408&8.142\\
\hline\hline
\end{tabular}
\end{table}

In this case, we fix the angular coverage in $360^{\circ}$ without position and sound velocity uncertainties. We define the different training environments with $8$, $16$, $32$, $56$ and $64$ detectors, respectively, and uniformly distributed in the angular coverage. Regarding testing environments, the lax environment is defined with $48$ detectors and the challenging one with only $4$. Notice that the restriction of 4 sensors in the challenging environment is a very harsh one and which is seldom to occur in practice. In fact, under such conditions, it is expected to have a very poor reconstruction performance, as the number of sensors is insufficient. However, we have chosen this challenging environment to see if robustness to this kind of invariance can be useful in such extreme situations.

The average results can be seen in Tab. \ref{tab:num_detectors}. While for the lax environment the best performance is achieved by the benchmark network (with the exception of PSNR), for the challenging environment ANDMask has a tiny advantage in some metrics. In any case, all the neural algorithms achieved a similar average performance. Rows $B1$ and $B2$ in Fig. \ref{fig:results_1} allow to appreciate some qualitative characteristics of the reconstructed images. It can be seen that, in the lax testing environment, the ANDMask trained network is capable of identifies some vessels and details present in the ground-truth image, but not recovered by the benchmark network. The performance of ANDMask and benchmark networks for the challenging environment is poor as expected.

\subsection{Results for environments with different angular coverage of the detection curve}

In this simulation, we fix the number of detectors in $32$ without position and sound velocity uncertainties. We define the training environments with an angular coverage of the detection curve of $120^{\circ}, 180^{\circ}, 240^{\circ}, 300^{\circ}$ and $360^{\circ}$, respectively. Regarding testing environments, the lax environment is defined with an angular coverage of $270^{\circ}$ and the challenging one with $60^{\circ}$. 
The result can be seen in Tab. \ref{tab:arc}. The model trained with ANDMask performs better than the benchmark network on all metrics except in the SSIM figure of merit of the lax environment. Again, $\tau=0.4$ seems to be the best hyperparameter for the lax environment and $\tau=0.8$ is the best for the challenging one. Fig. \ref{fig:results_2} (rows $C1$ and $C2$), show that for the lax environment, the ANDMask trained network presents a better reconstruction of some small vessels than the benchmark one. The reconstruction of those finer details is more regular and less choppy under the ANDMask network. For the challenging environment it is observed a slightly less blurred reconstruction of some of the structures of the ground-truth image. However, some artifacts, not present in the ground-truth image, are observed in both, the ANDMask and benchmark reconstructions.

\begin{table}
\caption{Algorithm performance against different angular coverage}
\label{tab:arc}
\setlength{\tabcolsep}{3pt}
\begin{tabular}{p{45pt}|p{70pt}|p{25pt}|p{23pt}|p{23pt}|p{23pt}|}
\hline\hline
Environment&Algorithm & SSIM & PC & RMSE & PSNR\\
\hline
Lax&ANDMask $\tau=0.8$&0.814&0.887&\bf{0.094}&\bf{21.039}\\
Lax&ANDMask $\tau=0.4$&0.813&\bf{0.895}&0.096&20.935\\
Lax&Benchmark&\bf{0.815}&0.891&0.095&20.974\\
Lax&LBP&-0.101&0.162&0.528&6.089\\\hline\hline
Challenging&ANDMask $\tau=0.8$&0.554&\bf{0.545}&0.232&13.022\\
Challenging&ANDMask $\tau=0.4$&\bf{0.566}&0.539&\bf{0.226}&\bf{13.272}\\
Challenging&Benchmark&0.558&0.535&0.232&12.991\\
Challenging&LBP&0.153&0.356&0.385&8.792\\
\hline\hline
\end{tabular}
\end{table}

\begin{table}
\caption{Algorithm performance against different sound velocity value uncertainties}
\label{tab:vel_uncertainties}
\setlength{\tabcolsep}{3pt}
\begin{tabular}{p{45pt}|p{70pt}|p{25pt}|p{23pt}|p{23pt}|p{23pt}|}
\hline\hline
Environment&Algorithm & SSIM & PC & RMSE & PSNR\\
\hline
Lax&ANDMask $\tau=0.8$&\bf{0.858}&\bf{0.960}&\bf{0.055}&\bf{26.130}\\
Lax&ANDMask $\tau=0.4$&0.854&0.958&0.056&25.913\\
Lax&Benchmark&\bf{0.858}&0.958&0.057&25.742\\
Lax&LBP&-0.117&0.243&0.534&6.024\\\hline\hline
Challenging&ANDMask $\tau=0.8$&0.634&0.803&0.119&19.202\\
Challenging&ANDMask $\tau=0.4$&\bf{0.636}&0.803&0.120&19.131\\
Challenging&Benchmark&0.643&\bf{0.814}&\bf{0.117}&\bf{19.257}\\
Challenging&LBP&0.021&0.211&0.403&8.104\\
\hline\hline
\end{tabular}
\end{table}

\subsection{Results for environments with different sound velocity value uncertainties}

In this experiment, we fix the number of detectors in $32$, the angular coverage in $360^{\circ}$ and without position uncertainty. We define the training environments  with sound velocity uncertainties of $0.01\,\%, 0.1\,\%, 0.5\,\%, 1.0\,\%$ and $1.5\,\%$, respectively. Regarding testing environments, the lax environment is defined with a sound velocity uncertainties of $0\,\%$ and the challenging one with $2\,\%$. The result can be seen in Tab. \ref{tab:vel_uncertainties}. These are the most surprising results, since the lax environment seems to be more sensitive to the use of the invariance through ANDMask than the challenging environment. However, it can also be interpreted (in the context of lack of data and spurious learning discussion of Section \ref{sec:ood}) that the error due to missing data is very sensitive to the uncertainty in speed. In this way, a variation in the uncertainty would affect this error more than the error due to spurious learning, making it more relevant not to discard information than to search for invariant representations. In any case, there are no big variations between the metrics with the exception of the challenging environment PC, where the benchmark network outperforms ANDMask but by a small margin. Finally, rows $D1$ and $D2$ in Fig. \ref{fig:results_2}  show some qualitative examples. For the lax environment,the ANDMask is able to perform a better reconstruction of finer details than the benchmark network. For the challenging environment, the quality of reconstruction is very similar for both the ANDMask and benchmark networks and, while some regions of the image seems to be more regular in the ANDMask reconstruction, some finer details are missing.

\section{Conclusions}
\label{sec:conclu}
We explored the application of invariance deep learning to the problem of OAT image reconstruction. In particular, we considered associated changes or uncertainties in the imaging setup with different environments on which the data collected is conditional. As an initial proof of concept we used the ANDMask training procedure developed in \cite{parascandolo21} and an appropriate neural network architecture (FD-UNet). Numerical experiments show that the neural networks trained using ANDMask presents in average a slightly  advantage with respect to similar neural network trained using a common and standard stochastic gradient descent scheme. From a purely qualitatively point of view, some examples were shown where the ANDMask trained model presented better quality reconstruction, in both lax and challenging environments. For example, the proposal were able to recover some finer details that the benchmark model was not able to recover. In other cases, there were no appreciable advantages. 

Although further studies might be necessary, the fact that the ANDMask is easily implementable, without significant extra computational effort, indicates that it could be a reasonable choice for training of invariant neural network models for OA imaging. As ANDMask and other invariant methods (such as IRM or IGA) were originally proposed for regression and classification problems, a future line of work could consider the study and  proposal of invariant methods specifically tailored for the problem of image reconstruction in OAT.



\subsection* {Acknowledgments}
This work was supported by the University of Buenos Aires (grant UBACYT 20020190100032BA), CONICET (grant PIP 11220200101826CO) and the ANPCyT (grants PICT 2018-04589, PICT 2020-01336).

\subsection* {Author's contributions}
All authors contributed equally to this work.

\subsection* {Data Availability Statement}
The data that support the findings of this study are available from the corresponding author upon reasonable request.



\bibliographystyle{ieeetr}

\bibliography{references}   


\end{document}